# A Fast Algorithm for the Analysis of Scattering by Elongated Cavities

Moti Zelig, Ehud Heyman, *Life Fellow, IEEE*, and Amir Boag, *Fellow, IEEE*

*Abstract*—The electromagnetic scattering from elongated, arbitrarily shaped, open-ended cavities have been studied extensively over the years. In this paper we introduce the fast encapsulating domain decomposition (EDD) scheme for the analysis of radar cross section (RCS) of such open-ended cavities. Problem definition, key principles, analysis, and implementation of the proposed solution scheme are presented in detail. The EDD advantages stem from domain decomposition along the elongated dimension and representing the fields on the cross-sections in the spectral domain, which enables us to separate the fields into in- and out-going waves. This diagonolizes the translation between the cross sections, thus reducing the per segment computational complexity from $O((N^A)^3)$ to $O(N^W(N^A)^2)$, where $N^A$ is the number of aperture unknowns and $N^W$ is the number of wall unknowns per segment, satisfying $N^W \ll N^A$, since we construct the segmentation step to be small compared to the cross section. The results of the EDD are demonstrated on an S-shape elongated open-ended cavity.

## LIST OF ABBREVIATIONS

| | |
|---|---|
| DD | domain decomposition |
| IE | integral equation |
| IO | input operator |
| CC | computational complexity |
| BEM | boundary elements method |
| CFIE | combined field IE |
| CIO | cavity IO |
| EDD | encapsulating DD |
| FEM | finite elements method |
| FFT | fast Fourier transform |
| FMM | fast multipole method |
| LCN | locally corrected Nyström |
| MoM | Method of Moments |
| NG | non-uniform grid |
| OEC | open-ended cavity |
| PEC | perfect electric conductor |
| RCS | radar cross section |
| RMS | root mean square |
| SBW | spectral bandwidth |
| SDD | segmented DD |
| SRR | spectral sampling rate |

## I. INTRODUCTION

THE radar cross section (RCS) of elongated open-ended cavity (OEC), such as a jet-engine inlet, provides a rich signature that can be used for target classification and identification. The large electrical dimensions, non-canonical shape, and internal resonances make it extremely difficult to calculate correctly the fields and currents excited by the incident wave in a typical cavity. Existing computational schemes, such as, modal expansions, ray/beam approximations, and integral- and differential- equation based solvers, suffer from a mixture of drawbacks [1]: including difficulties in handling arbitrary geometries, limited accuracy and stability, high computational complexity, and, at times, chaotic behavior.

It has been proposed to overcome these obstacles for scatterers involving OECs by exploiting the Love's equivalence principle [2] to perform various domain decomposition (DD) schemes [3]–[11]. In basic DD, the problem in hand is divided into two regions, the inner problem, i.e. the OEC and the outer problem of the enclosing shell. Highly elongated OEC is often further subdivided into segments along its long axis. The idea is to march from the close-end to the open-end and calculate the input operator (IO) of each segment, taking into account the previously calculated segment IO which can be considered as a load for the current segment. The algorithm guarantees the fields' continuity on the coupling (common) interface. The cavity IO (CIO) is used as a generalized impedance boundary condition in the solution scheme for the outer problem. By using a DD based scheme, the number of unknowns to deal with, at each step is much lower than when using a brute-force scheme, hence, improved numerical efficiency is achieved. The overall computational complexity of the segmented DD is of $O(L(N^A)^3)$ assuming that $N^W \ll N^A$, where $N^A$ is the number of aperture (coupling interface) unknowns of the segment, $N^W$ is the number of wall unknowns per segment, and $L$ is the number of segments. The computational complexity of calculating the OEC as a whole is of $O((N^C)^3)$. It is not guaranteed that the segmented DD is more efficient than solving the OEC without segmentation. The efficiency depends on the ratio between the OEC's wall area and the aperture's area.

The proposed solution scheme, the fast Encapsulating Domain Decomposition (EDD) scheme, which preliminary results have been reported in conference papers [12], [13], is based on representing the coupling operator in the spectral domain, which has two main advantages: the translation from one coupling interface to the next one is diagonal and it enables us to apply directional decomposition of the fields on the coupling interfaces. These properties reduce the per segment computational complexity from $O((N^A)^3)$ to $O(N^W(N^S)^2)$, where $N^S$ is the number of spectral unknowns, $O(N^S) = O(N^A)$. The overall computational complexity of analyzing the OEC, using the EDD is of $O(LN^W(N^S)^2)$, which is guaranteed to be more efficient than the segmented DD solution of the OEC as a whole.

The outline of this paper is as follows. Section II presents the problem description, Section III - the EDD, while Section IV is devoted to the details of the spatial and spectral discretization. In Section V, the efficiency and accuracy of the EDD scheme using an S-shape OEC enclosed by an external shell, are demonstrated. The numerical results are then compared to those of a reference solution, a brute-force combined field integral equation (CFIE) with a locally corrected Nyström (LCN) discretization, basic DD scheme, separated into outer and inner problems (DD$_0$-LCN) and segmented DD-LCN. Finally, the conclusions and summary are presented in Section VI.

## II. Problem Description

Consider the problem of electromagnetic scattering from an object comprising an open-ended cavity (OEC) enclosed in a shell as depicted in Fig. 1(a). A harmonic time dependence $\exp(j\omega t)$ is assumed and suppressed. The proposed algorithm is based on the frequency domain integral equation approach. As a background, this section reviews the existing solution schemes, beginning with the description of a global solution of the problem at hand, then turning to the basic domain decomposition (DD) into inner and outer problems, and ending with the segmented DD (SDD).

### A. Global Integral Equation Solution Scheme

We employ the integral equation (IE) solution approach, which is based on the integral representation of the fields:

$$\mathbf{E}^i = \mathbf{E} + \eta \mathcal{L}(\mathbf{J}) + \mathcal{K}(\mathbf{M}) \qquad (1)$$

and

$$\mathbf{H}^i = \mathbf{H} - \mathcal{K}(\mathbf{J}) + \eta^{-1} \mathcal{L}(\mathbf{M}), \qquad (2)$$

where $\mathbf{E}$ and $\mathbf{H}$ are the total electric and magnetic fields, respectively, $\mathbf{E}^i$ and $\mathbf{H}^i$ are the incident fields, $\mathbf{J}$ and $\mathbf{M}$ are the electric and magnetic current densities (physical or equivalent), $\eta$ is the free space impedance, and operators $\mathcal{L}$ and $\mathcal{K}$ (used in [5], [6], [14]–[16]) are defined as

$$\left\{ \begin{matrix} \mathcal{L} \\ \mathcal{K} \end{matrix} \right\}(\mathbf{F}) = \left\{ \begin{matrix} jk^{-1}(k^2 + \nabla\nabla \cdot) \\ \nabla \times \end{matrix} \right\} \int \mathbf{F}(\mathbf{r}') G(\mathbf{r},\mathbf{r}') d\mathbf{r}' \qquad (3)$$

where $\mathbf{F}$ is either $\mathbf{J}$ or $\mathbf{M}$, $\mathbf{r}$ and $\mathbf{r}'$ are the observation and source points, respectively, and $G(\mathbf{r},\mathbf{r}') = (4\pi|\mathbf{r}-\mathbf{r}'|)^{-1} e^{-jk|\mathbf{r}-\mathbf{r}'|}$ is the free-space Green's function with $k$ being the wavenumber. The scattering problem may be addressed now by

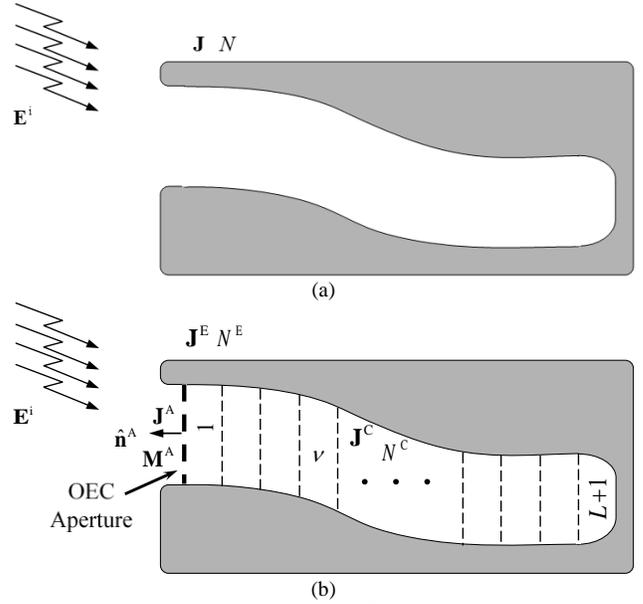

Fig. 1. (a) Radar target comprising an OEC. (b) Problem decomposition to inner and outer problems with OEC domain decomposition by segmentation.

replacing the target with a surface whereon we set Love's equivalent currents [2] $\mathbf{J} = \hat{\mathbf{n}} \times \mathbf{H}$ and $\mathbf{M} = -\hat{\mathbf{n}} \times \mathbf{E}$, where $\hat{\mathbf{n}}$ is the outward normal to the surface. For a PEC target, the total tangential electric field vanishes on the surface, hence by applying (1)-(2) at all points $\mathbf{r}$ on the surface, we obtain the IE set

$$\hat{\mathbf{n}} \times \mathbf{E}^i = \eta \hat{\mathbf{n}} \times \mathcal{L}(\mathbf{J}) \qquad (4)$$

and

$$\hat{\mathbf{n}} \times \mathbf{H}^i = -\hat{\mathbf{n}} \times \left(\tfrac{1}{2}\mathcal{I} + \hat{\mathcal{K}}\right)(\mathbf{J}) \qquad (5)$$

where $\mathcal{I}(\mathbf{F}) = \hat{\mathbf{n}} \times \mathbf{F}$ is a rotation operator on the surface. Note that since the points $\mathbf{r}$ are located now on the surface, we used the limit

$$\mathcal{K}(\mathbf{F}) \to -\tfrac{1}{2}\hat{\mathbf{n}} \times \mathbf{F} + \hat{\mathcal{K}}(\mathbf{F}) \qquad (6)$$

where the integral operator $\hat{\mathcal{K}}$ is defined in the principal value sense. Each of the IE's in (4) and (5) may be solved in isolation, yet, in order to eliminate possible spurious effects at the internal resonance frequencies, we utilize them in a CFIE setup [17].

The computational complexity of solving this problem directly, using BEM, is of $O(N^3)$, and the required storage space is of $O(N^2)$ with $N = N^E + N^C$ where $N^E$ and $N^C$ denote the number of unknowns on the external surface and in the cavity. For realistic targets at typical radar frequencies, this overall number of unknown $N$ is extremely large. Such large problems are typically solved using fast iterative algorithms with pre-conditioners. However, due to the strongly resonant nature of the internal interactions in the cavity, the impedance matrix describing these interactions is expected to be highly ill-conditioned, thus making such iterative schemes converge very slowly, if at all. It has been proposed to overcome this difficulty by separating the problem into inner and outer ones, as described in the following subsection.

*B. Basic Domain Decomposition: Separation into Inner and Outer Problems*

The *inherent numerical difficulty*, which stems from different physical nature of the internal and external scattering mechanisms, may be addressed by separating the inner and outer problems, and solving the latter by considering the former as a loading. Each domain may then be solved by a different numerical approach that best fits its characteristics.

This DD approach has another practical advantage. In many applications, the external structure of the target may be modified for various reasons, such as design considerations, while the cavity structure is usually unchanged. In the DD approach, the cavity problem is solved once and then used for different realizations of the external configuration.

*1) The cavity input impedance operator (CIO).*

The inherent difficulty mentioned above is addressed here by separating the inner and outer problems, as illustrated in Fig. 1(b), and solving the latter by considering the former as a loading. By the equivalence principle [2], [18], the effect of the cavity on the external problem is fully described by the equivalent currents $\mathbf{J}^A = \hat{\mathbf{n}}^A \times \mathbf{H}$ and $\mathbf{M}^A = -\hat{\mathbf{n}}^A \times \mathbf{E}$ at the cavity's aperture, where $\hat{\mathbf{n}}^A$ is the outward normal to the aperture. The boundary conditions inside the cavity impose a relation between $\mathbf{J}^A$ and $\mathbf{M}^A$ that can be expressed in terms of the CIO $\mathscr{Z}^A$ via

$$\mathbf{M}^A = \mathscr{Z}^A \mathbf{J}^A, \qquad (7)$$

The CIO is found by calculating $\mathbf{M}^A$ for an appropriate set of excitations $\{\mathbf{J}_n^A\}$ that span the space of possible current distributions $\mathbf{J}^A$. These calculations can be performed via various methods, e.g., finite elements method (FEM) [3] or BEM [4], where the choice typically depends on the properties of the cavity. Here, we follow the IE approach based on (1). Inside the PEC cavity, (1) becomes $0 = \mathbf{E} + \eta \mathcal{L}(-\mathbf{J}^A) + \eta \mathcal{L}(\mathbf{J}^C) + \mathcal{K}(-\mathbf{M}^A)$ where $\mathbf{J}^C$ is the physical current on the internal cavity walls, the minus signs of $\mathbf{J}^A$ and $\mathbf{M}^A$ are due to the fact that these equivalent currents were defined above with respect to the outward normal $\hat{\mathbf{n}}^A$. The IE for $\mathbf{J}^C$, $\mathbf{J}^A$ and $\mathbf{M}^A$ is obtained now by letting the observation point, $\mathbf{r}$, approach the domain boundaries, i.e., the aperture or the cavity walls. Now, (1) can be cast in the following format

$$\begin{pmatrix} -\mathbf{Z}^{AA} & -\mathbf{T}^{AA} & \mathbf{Z}^{AC} \\ -\mathbf{Z}^{CA} & -\mathbf{T}^{CA} & \mathbf{Z}^{CC} \end{pmatrix} \begin{pmatrix} \mathbf{J}^A \\ \mathbf{M}^A \\ \mathbf{J}^C \end{pmatrix} = \begin{pmatrix} 0 \\ 0 \end{pmatrix} \qquad (8)$$

where $\mathbf{Z}^{OS} = \eta \hat{\mathbf{n}}^O \times \mathcal{L}^{OS}$ and $\mathbf{T}^{OS} = \hat{\mathbf{n}}^O \times (\tfrac{1}{2}\mathcal{I} + \widehat{\mathcal{K}}^{OS})$, and we use the superscript notation $O$ and $S$ for the observation and source domains, which can be either "A" for the aperture or "C" for cavity walls. Note that the rotation operator $\mathcal{I}$ is included only in the $\mathbf{T}^{AA}$ term in (8), and that the rotation is defined as in (5) with respect to the normal that points inward toward the cavity interior. The CIO in (7) is found by applying the Schur's complement procedure to (8), thus obtaining

$$\mathscr{Z}^A = -\left(\mathbf{T}^{AA} - \mathbf{Z}^{AC}\left(\mathbf{Z}^{CC}\right)^{-1}\mathbf{T}^{CA}\right)^{-1} \\ \left(\mathbf{Z}^{AA} - \mathbf{Z}^{AC}\left(\mathbf{Z}^{CC}\right)^{-1}\mathbf{Z}^{CA}\right). \qquad (9)$$

We refer to this solution as $DD_0$ (basic domain decomposition, in order to distinguish it from the segmented domain decomposition (SDD) or the encapsulated domain decomposition (EDD) to be discussed below. The computational complexity of calculating the CIO via (9), taking into account all the dominant matrix operations, is given by

$$CC^{DD_0} \approx \tfrac{2}{3}(N^C)^3 + 4N^A(N^C)^2 + 4(N^A)^2 N^C + \tfrac{4}{3}(N^A)^3, \qquad (10)$$

The storage needed for the preprocessed operators is given by

$$Strg^{DD_0} = 2(N^A)^2 + 3N^A N^C + (N^C)^2 \qquad (11)$$

*2) Solution of the external scattering problem*

The integral equations for the external domain obtained from (1) and (2) are given therefore by the CFIE set

$$\begin{pmatrix} \mathbf{Z}^{EE} & \mathbf{Z}^{EA} & \mathbf{T}^{EA} \\ \mathbf{Z}^{AE} & \mathbf{Z}^{AA} & \mathbf{T}^{AA} \end{pmatrix} \begin{pmatrix} \mathbf{J}^E \\ \mathbf{J}^A \\ \mathbf{M}^A \end{pmatrix} = \begin{pmatrix} \hat{\mathbf{n}}^E \times \mathbf{E}^{i,E} \\ \hat{\mathbf{n}}^A \times \mathbf{E}^{i,A} \end{pmatrix} \qquad (12)$$

and

$$\begin{pmatrix} -\mathbf{T}^{EE} & -\mathbf{T}^{EA} & \mathbf{Y}^{EA} \\ -\mathbf{T}^{AE} & -\mathbf{T}^{AA} & \mathbf{Y}^{AA} \end{pmatrix} \begin{pmatrix} \mathbf{J}^E \\ \mathbf{J}^A \\ \mathbf{M}^A \end{pmatrix} = \begin{pmatrix} \hat{\mathbf{n}}^E \times \mathbf{H}^{i,E} \\ \hat{\mathbf{n}}^A \times \mathbf{H}^{i,A} \end{pmatrix} \qquad (13)$$

where $\mathbf{Y}^{OS} = \eta^{-1}\hat{\mathbf{n}}^O \times \mathcal{L}^{OS}$ and we use the superscript notation defined in (8) with superscript "E" denoting the external surface domain. Likewise, $\mathbf{J}^E$ is the physical electric current on the external surface while $\mathbf{E}^{i,E}$, $\mathbf{H}^{i,E}$, $\mathbf{E}^{i,A}$, and $\mathbf{H}^{i,A}$ are the incident electrical and magnetic fields on the external walls and on the aperture, while $\hat{\mathbf{n}}^E$ is outward normal at the external walls. Since $\mathbf{J}^A$ and $\mathbf{M}^A$ are related via the CIO (7), we can cast (12)-(13) in the form

$$\begin{pmatrix} \mathbf{Z}^{EE} & \mathbf{Z}^{EA} + \mathbf{T}^{EA}\mathscr{Z}^A \\ \mathbf{Z}^{AE} & \mathbf{Z}^{AA} + \mathbf{T}^{AA}\mathscr{Z}^A \end{pmatrix} \begin{pmatrix} \mathbf{J}^E \\ \mathbf{J}^A \end{pmatrix} = \begin{pmatrix} \hat{\mathbf{n}}^E \times \mathbf{E}^{i,E} \\ \hat{\mathbf{n}}^A \times \mathbf{E}^{i,A} \end{pmatrix} \qquad (14)$$

and

$$\begin{pmatrix} -\mathbf{T}^{EE} & -\mathbf{T}^{EA} + \mathbf{Y}^{EA}\mathscr{Z}^A \\ -\mathbf{T}^{AE} & -\mathbf{T}^{AA} + \mathbf{Y}^{AA}\mathscr{Z}^A \end{pmatrix} \begin{pmatrix} \mathbf{J}^E \\ \mathbf{J}^A \end{pmatrix} = \begin{pmatrix} \hat{\mathbf{n}}^E \times \mathbf{H}^{i,E} \\ \hat{\mathbf{n}}^A \times \mathbf{H}^{i,A} \end{pmatrix} \qquad (15)$$

Finally, the scattered electric field is obtained by substituting the surface currents $\mathbf{J}^E$, $\mathbf{J}^A$, and $\mathbf{M}^A$, obtained by utilizing CFIE solution combining (14) and (15), into (1).

The basic domain decomposition yields several favorable properties, among them are the flexibly to use the most suitable solution scheme for each sub-problem, and reduced computational complexity and storage requirements which stem from treating the inner and outer problems separately.

The present work is not concerned with the solution of the outer problem. We shall only comment that the complexity of a brute-force solution of this problem is of $O((N^E)^3)$, but by using fast iterative schemes, such as the FMM [19], [20] or NG algorithm [21]–[23], it can be reduced to $O(N^E \log N^E)$. The overall computational complexity is given, therefore, by this term plus $CC^{DD_0}$ of (10), which is essentially $O((N^C)^3)$. Thus, the computational complexity of the DD-based solution is lower than that of the brute-force global direct solution which, as noted after (6), is $O((N^E + N^C)^3)$.

### C. Segmented Domain Decomposition for the Inner Problem

In the previous section, the problem of calculating the CIO has been formulated as a single integral equation over the cavity's boundaries (see (8)). As noted there, the resonative nature of the interactions inside the cavity prevents a fast iterative solution, and one should resorts to the brute force inversion in (9), whose computational complexity is of $O(N^C)^3$ see (10).

For an elongated cavity as in Fig. 1, an alternative approach is to subdivide the cavity into segments along the cavity axis, as in Fig. 1(b), and then calculate the input operator (IO) of each segment in terms of the IO of next segment. The IO of the entire cavity is then obtained by repeating this procedure recursively. This solution approach, first presented in [4], is more stable than the global approach discussed above since the sections are less resonative than the entire cavity.

Referring to Fig. 1(b), the cavity is subdivided into $L+1$ segments, tagged by the index $\nu$, where $\nu = 1$ designates the open-end segment and $L+1$ is the termination. In many cases, the termination segment is quite complicated, hence its IO may be calculated separately by an appropriate scheme, e.g. using FEM. Fig. 2 depicts a typical segment $\nu$, $1 \le \nu \le L$, whose interfaces with the neighboring $\nu - 1$ and $\nu + 1$ segments are denoted as "Input" and "Load", respectively. The unknown fields at these interfaces are described by the equivalent currents. Specifically, the unknown at the interface between the segments $\nu$ and $\nu + 1$ are

$$\mathbf{J}_\nu^L = \hat{\mathbf{n}}_\nu^L \times \mathbf{H}, \qquad \mathbf{M}_\nu^L = -\hat{\mathbf{n}}_\nu^L \times \mathbf{E} \qquad (16)$$

where $\hat{\mathbf{n}}_\nu^L$ is the normal pointing toward the open end of the cavity. These unknown currents are related via an IE whose derivation is similar to that of (8). Considering an observation point inside the $\nu$th segment, (1) becomes

$$0 = \mathbf{E} + \eta \mathcal{L}(-\mathbf{J}_{\nu-1}^L) + \mathcal{K}(-\mathbf{M}_{\nu-1}^L) \\ + \eta \mathcal{L}(\mathbf{J}_\nu^L) + \mathcal{K}(\mathbf{M}_\nu^L) + \eta \mathcal{L}(\mathbf{J}_{\nu-1}^W) \qquad (17)$$

where $\mathbf{J}^W$ is the physical current on the internal walls. The IE is obtained now by letting $\mathbf{r}$ approach the segment boundaries. It can be cast in the following format (see (8)):

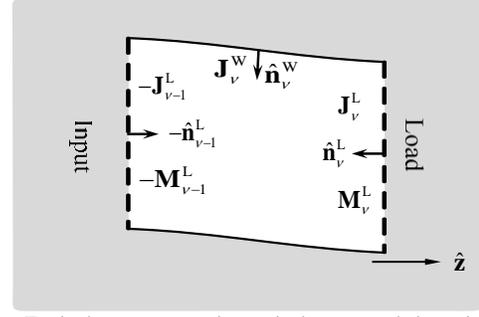

Fig. 2. Equivalent currents in typical open-ended cavity domain decomposition segment.

$$\begin{pmatrix} -\mathbf{Z}_\nu^{II} & -\mathbf{T}_\nu^{II} & \mathbf{Z}_\nu^{IW} & \mathbf{Z}_\nu^{IL} & \mathbf{T}_\nu^{IL} \\ -\mathbf{Z}_\nu^{WI} & -\mathbf{T}_\nu^{WI} & \mathbf{Z}_\nu^{WW} & \mathbf{Z}_\nu^{WL} & \mathbf{T}_\nu^{WL} \\ -\mathbf{Z}_\nu^{LI} & -\mathbf{T}_\nu^{LI} & \mathbf{Z}_\nu^{LW} & \mathbf{Z}_\nu^{LL} & \mathbf{T}_\nu^{LL} \end{pmatrix} \begin{pmatrix} \mathbf{J}_{\nu-1}^L \\ \mathbf{M}_{\nu-1}^L \\ \mathbf{J}_\nu^W \\ \mathbf{J}_\nu^L \\ \mathbf{M}_\nu^L \end{pmatrix} = \begin{pmatrix} 0 \\ 0 \\ 0 \end{pmatrix}, \qquad (18)$$

where the operators $\mathbf{Z}_\nu^{OS}$ and $\mathbf{T}_\nu^{OS}$ have been defined in (8) and we use the same superscript notation $O$ and $S$ for the observation and source domains, which can be "L", "I" and "W" for the Load interface, the Input interface and the cavity-Walls, respectively. In the recursive solution process, the effect of successive segments is inserted into (18) in the form of the IO at the Load interface,

$$\mathbf{M}_\nu^L = \mathcal{Z}_\nu^L \mathbf{J}_\nu^L. \qquad (19)$$

The IO of the $\nu^{\text{th}}$ segment, $\mathbf{M}_{\nu-1}^L = \mathcal{Z}_{\nu-1}^L \mathbf{J}_{\nu-1}^L$, is then found from (18) and (19). An explicit expression of this relation is given in Appendix A. Finally, the CIO is given by $\mathcal{Z}_0^L$, i.e., the IO of the $\nu = 1$ segment.

As noted above, the termination segment is often quite complicated, hence its IO $\mathcal{Z}_L^L$ is calculated separately by an appropriate scheme. If, however, it is composed of simple PEC walls, then its IO may be calculated via the IE approach, presented here, giving (see (9))

$$\mathcal{Z}_L^L = -\left( \mathbf{T}_{L+1}^{II} - \mathbf{Z}_{L+1}^{IW} \left( \mathbf{Z}_{L+1}^{WW} \right)^{-1} \mathbf{T}_{L+1}^{WI} \right)^{-1} \\ \left( \mathbf{Z}_{L+1}^{II} - \mathbf{Z}_{L+1}^{IW} \left( \mathbf{Z}_{L+1}^{WW} \right)^{-1} \mathbf{Z}_{L+1}^{WI} \right). \qquad (20)$$

In the SDD, the above operators stem from convolution integrals, hence, resulting in fully populated matrices. The computational cost of finding the IO of the segment is of $O((N^A)^3)$, where $N^A$ is the number of interface unknowns. To this end, the overall computational complexity, $CC^{SDD}$, of calculating the CIO via (18), excluding second order operation such as matrix summations is given by (see Appendix A)

$$CC^{SDD} \approx \\ \frac{2}{3} L^{-2} (\hat{N}^C)^3 + 6 L^{-1} N^A (\hat{N}^C)^2 + 14 (N^A)^2 \hat{N}^C + \frac{46}{3} L (N^A)^3 \qquad (21)$$

and the optimal number of segments is given by

$$L_{\text{Opt}}^{SDD} = \sqrt{18/46} (\hat{N}^C / N^A) \qquad (22)$$

where $\hat{N}^C$ is the number of elements in the cavity, excluding those in the termination segment. The storage $Strg^{SDD}$ needed for the preprocessed operators of a single segment is given by

$$Strg^{SDD} = 9(N^A)^2 + L^{-2}(\widehat{N}^C)^2 + 6L^{-1}N^A\widehat{N}^C \qquad (23)$$

One readily discerns from (21) that for small $L$, $CC^{SDD}$ is governed by $\widehat{N}^C$, while for large $L$ it is dominated by $L(N^A)^3$, representing the inversions and multiplications of the operators at all the interfaces between segments.

Fig. 3 depicts a parametric comparison of the potential speedup and storage reduction in the SDD vs. the DD$_0$ as a function of the number of segments $L$, the ratio $N^A/N^C$, and the number of elements in the termination segment $N^T/N^C$. The SDD is advantageous in the parameter range depicted by the shaded area. The termination has a very little influence on the efficiency. The optimal choice of the number of segments, given in (22), as long as the number of segments multiplied by the inlet cross section number of unknowns is less than the wall one, which result in quite coarse segmentation, and for wide OEC the SDD efficiency is similar to the DD$_0$.

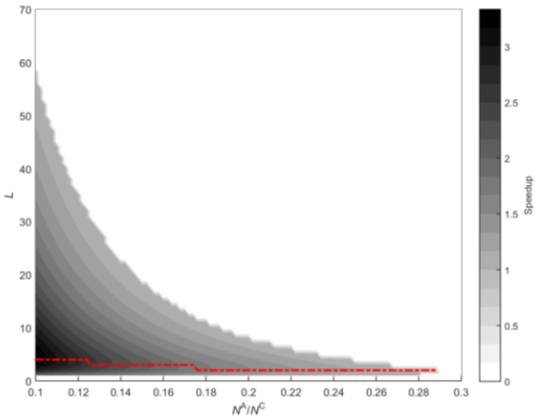

Fig. 3. SDD vs. DD$_0$ potential Speedup as a function of number of segments $L$ and $N^A/N^C$. The potential speedup is indicated by the gray shade (see (21)) while the red dash-dot line delineates the speedup with optimal segmentation (see (22)). The optimal speedup is about 2.5 and it is obtained for small number of segments.

To this end, one can readily observes that the SDD speedup is getting saturated very quickly. Using a finer segmentation of the inlet makes the solution inefficient. In the following section, we present our novel efficient approach, the EDD.

### III. ENCAPSULATING DOMAIN DECOMPOSITION

In this work, we propose a novel approach, the EDD, based on the SDD scheme, presented in Section II.C, which is solved by using a spectral-spatial formulation. The spectral formulation separates between forward and backward propagating waves and therefore simplifies the recursive solution of the CIO. As a result, the computation complexity of the spatial and spectral formulations are different, hence these alternative formulations constitute a tradeoff where the choice depends on the physical problem. In this section, we first establish the spectral representation, then we utilize it in the EDD, our solution scheme, and the section ends with computational complexity analysis.

#### A. Spectral representation of the field

In general, the segmentation along the cavity axis used in Section II.C may be quite general. Referring to Fig. 1, the segmentation here is such that all the interfaces are planar and are normal to the longitudinal direction of the overall cavity, henceforth denoted as the $z$ axis. For brevity, the spatial and spectral transversal coordinates are denoted, respectively, as $\mathbf{x}_t = (x, y)$ and $\mathbf{k}_t = (k_x, k_y)$. The spectral representation of any field constituent $\mathbf{A}$ is marked by a tilde and defined as

$$\tilde{\mathbf{A}}(\mathbf{k}_t) = \mathcal{F}\{\mathbf{A}(\mathbf{x}_t)\} = \frac{1}{2\pi}\int d^2 x_t e^{j\mathbf{k}_t \cdot \mathbf{x}_t}\mathbf{A}(\mathbf{x}_t), \qquad (24)$$

such that

$$\mathbf{A}(\mathbf{x}_t) = \mathcal{F}^{-1}\{\tilde{\mathbf{A}}(\mathbf{k}_t)\} = \frac{1}{2\pi}\int d^2 k_t e^{-j\mathbf{k}_t \cdot \mathbf{x}_t}\tilde{\mathbf{A}}(\mathbf{k}_t). \qquad (25)$$

We now consider the radiation of any *transversal* sources $(\mathbf{J}, \mathbf{M})$ residing on a given $z'$ plane. Denoting the spectral form of these sources as $\tilde{\mathbf{F}}(\mathbf{x}_t)$, the spectral representation of the surface integral operators in (3) is given by

$$\begin{Bmatrix}\tilde{\mathcal{L}}\\ \tilde{\mathcal{K}}\end{Bmatrix}(\tilde{\mathbf{F}}) = \begin{Bmatrix}-jk\hat{\mathbf{k}}^{\pm} \times \hat{\mathbf{k}}^{\pm} \times \\ -jk\hat{\mathbf{k}}^{\pm} \times \end{Bmatrix}\tilde{\mathbf{F}}(\mathbf{k}_t; z')\tilde{g}(\mathbf{k}_t; z, z') \qquad (26)$$

where

$$\tilde{g}(\mathbf{k}_t; z, z') = \frac{e^{-jk_z|z'-z|}}{2jk_z}, \qquad (27)$$

is the spectral 1D Green's function, with

$$k_z = \sqrt{k^2 - k_t^2}, \quad \operatorname{Im}\{k_z\} \le 0, \qquad (28)$$

and

$$\mathbf{k}^{\pm} = \begin{cases}(\mathbf{k}_t, k_z) & z > z' \\ (\mathbf{k}_t, -k_z) & z < z'\end{cases}, \quad \hat{\mathbf{k}}^{\pm} = \mathbf{k}^{\pm}/k, \qquad (29)$$

where $\mathbf{k}^{\pm}$ are the forward/backward propagating 3D wave-vectors associated with $\mathbf{k}_t$. Substituting (26) into (1), the spectral electric field at a given plane $z > z'$ or $z < z'$ is given by

$$\tilde{\mathbf{E}} = jk\tilde{g}(\mathbf{k}_t; z, z')\left(\eta\hat{\mathbf{k}}^{\pm} \times \hat{\mathbf{k}}^{\pm} \times \tilde{\mathbf{J}}(\mathbf{k}_t; z') + \hat{\mathbf{k}}^{\pm} \times \tilde{\mathbf{M}}(\mathbf{k}_t; z')\right) (30)$$

Expression (30) can be cast in the form

$$\tilde{\mathbf{E}} = \tilde{S}(\mathbf{k}_t; z, z')\begin{cases}\tilde{\mathbf{V}}^+ & z > z' \\ \tilde{\mathbf{V}}^- & z < z'\end{cases}, \qquad (31)$$

where

$$\tilde{S}(\mathbf{k}_t; z, z') = e^{-jk_z|z'-z|}. \qquad (32)$$

is the spectral propagator, and

$$\begin{pmatrix}\tilde{\mathbf{V}}^+ \\ \tilde{\mathbf{V}}^-\end{pmatrix} = \frac{jk}{2jk_z}\begin{pmatrix}\eta\hat{\mathbf{k}}^+ \times \hat{\mathbf{k}}^+ \times & \hat{\mathbf{k}}^+ \times \\ \eta\hat{\mathbf{k}}^- \times \hat{\mathbf{k}}^- \times & \hat{\mathbf{k}}^- \times\end{pmatrix}\begin{pmatrix}\tilde{\mathbf{J}} \\ \tilde{\mathbf{M}}\end{pmatrix}. \qquad (33)$$

are the vector-amplitudes of the forward and backward propagating plane waves. Note that $\tilde{\mathbf{V}}^{\pm}$ are, in fact, fully described by their transversal components $\tilde{\mathbf{U}}^{\pm} = -\hat{\mathbf{z}} \times \hat{\mathbf{z}} \times \tilde{\mathbf{V}}^{\pm}$, such that by using the orthogonality condition $\mathbf{k}^{\pm} \cdot \tilde{\mathbf{V}}^{\pm} = 0$ we have

$$\tilde{\mathbf{V}}^{\pm} = \mathbf{R}_z^{\pm} \tilde{\mathbf{U}}^{\pm}, \quad \mathbf{R}_z^{\pm} = \mathbf{I} \mp \hat{\mathbf{z}}(k_z)^{-1} \mathbf{k} \cdot, \quad (34)$$

where $\mathbf{I}$ is the identity operator. $\mathbf{R}_z^{\pm}$ will be referred to as the plane-wave reconstruction operators.

Finally, we note that if $(\mathbf{J}, \mathbf{M})$ are the equivalent currents as defined in (16) (recall that $\hat{\mathbf{n}}_v^L = -\hat{\mathbf{z}}$ here), then

$$\begin{pmatrix} \tilde{\mathbf{J}} \\ \tilde{\mathbf{M}} \end{pmatrix} = \begin{pmatrix} -\eta^{-1} \hat{\mathbf{z}} \times \hat{\mathbf{k}}^+ \times & -\eta^{-1} \hat{\mathbf{z}} \times \hat{\mathbf{k}}^- \times \\ \hat{\mathbf{z}} \times & \hat{\mathbf{z}} \times \end{pmatrix} \begin{pmatrix} \tilde{\mathbf{V}}^+ \\ \tilde{\mathbf{V}}^- \end{pmatrix}. \quad (35)$$

*B. The EDD Solution Scheme*

We now apply the spectral formulation to the recursive DD formulation of the CIO. We recall that the DD segmentation has been discussed at the beginning of Section III.A, thus, by transforming rows 1 and 3 in the integral equation (18) of segment $v$ to the spectral domain and using (35), we obtain

$$\begin{pmatrix} 0 & -\mathcal{I}_v^I \mathcal{F} \mathbf{Z}_v^{IW} & \tilde{S}_v^{IL} \\ -\mathcal{F}^{-1} \mathcal{I}_v^W \tilde{S}_v^{WI} \mathbf{R}_z^+ & \mathbf{Z}_v^{WW} & \mathcal{F}^{-1} \mathcal{I}_v^W \tilde{S}_v^{WL} \mathbf{R}_z^- \\ \tilde{S}_v^{LI} & \mathcal{I}_v^L \mathcal{F} \mathbf{Z}_v^{LW} & 0 \end{pmatrix} \begin{pmatrix} \tilde{\mathbf{U}}_{v-1}^+ \\ \mathbf{J}_v^W \\ \tilde{\mathbf{U}}_v^- \end{pmatrix}$$
$$= \begin{pmatrix} \tilde{\mathbf{U}}_{v-1}^- \\ 0 \\ \tilde{\mathbf{U}}_v^+ \end{pmatrix} \quad (36)$$

where $\tilde{\mathbf{U}}_{v-1}^{\pm}$, and $\tilde{\mathbf{U}}_v^{\pm}$ are the forward/backward propagating transverse plane-wave amplitudes at the Input and Load interfaces of segment $v$, $\mathcal{I}_v^O = \hat{\mathbf{n}}_v^O \times$ rotation operator, defined in (5), $\mathbf{R}_z^{\pm}$ is the plane wave $z$ component reconstruction operator, defined in (34), and we utilize the notation introduced in (8) and (18). Note that the Fourier operator $\mathcal{F}$ in the first and third rows is performed over input and load interfaces, respectively, whereas the $\mathcal{F}^{-1}$ operator in the second row is performed for observation points on the walls at any $z$. It is therefore preferable to execute $\mathcal{F}^{-1}$ via direct integration of (25) and not via the FFT algorithm. The recursive solution for the CIO starts from the termination segment and proceeds toward the input, where each segment acts as a load (an IO) for the preceding segment. While in the formulation of (18), this IO is an impedance (or admittance) operators, in (36), the IO is a reflection operator, transforming a spectrum of plane waves propagating into the interface (toward the termination) into a spectrum of plane waves propagating toward the open-end. The spectral domain representation of this operator is

$$\tilde{\mathbf{U}}_v^-(\mathbf{k}_t) = \tilde{\mathbf{\Gamma}}_v^L(\mathbf{k}_t, \mathbf{k}_t') \tilde{\mathbf{U}}_v^+(\mathbf{k}_t'). \quad (37)$$

Eq. (36) is solved therefore by substituting (37), for $\tilde{\mathbf{U}}_v^-$ and then eliminating $\tilde{\mathbf{U}}_v^+$ via a Schur's complement procedure. A convenient form to solve this equation is via the signal flow graph representation of (36) depicted in Fig. 4(a). Analyzing this graph, the input reflection operator is found to be

$$\tilde{\mathbf{\Gamma}}_{v-1}^L = \tilde{S}_v^{IL} \tilde{\mathbf{\Gamma}}_v^L \tilde{S}_v^{LI} - \\ \left[ -\mathcal{I}_v^I \mathcal{F} \mathbf{Z}_v^{IW} + \tilde{S}_v^{IL} \tilde{\mathbf{\Gamma}}_v^L \mathcal{I}_v^L \mathcal{F} \mathbf{Z}_v^{LW} \right] \left( \mathbf{Z}_v^{WW} \right)^{-1} \\ \times \left[ I + \mathcal{F}^{-1} \mathcal{I}_v^W \tilde{S}_v^{WL} \mathbf{R}_z^- \tilde{\mathbf{\Gamma}}_v^L \mathcal{I}_v^L \mathcal{F} \mathbf{Z}_v^{LW} \left( \mathbf{Z}_v^{WW} \right)^{-1} \right]^{-1} \quad (38) \\ \times \left[ -\mathcal{F}^{-1} \mathcal{I}_v^W \tilde{S}_v^{WI} \mathbf{R}_z^+ + \mathcal{F}^{-1} \mathcal{I}_v^W \tilde{S}_v^{WL} \mathbf{R}_z^- \tilde{\mathbf{\Gamma}}_v^L \tilde{S}_v^{LI} \right].$$

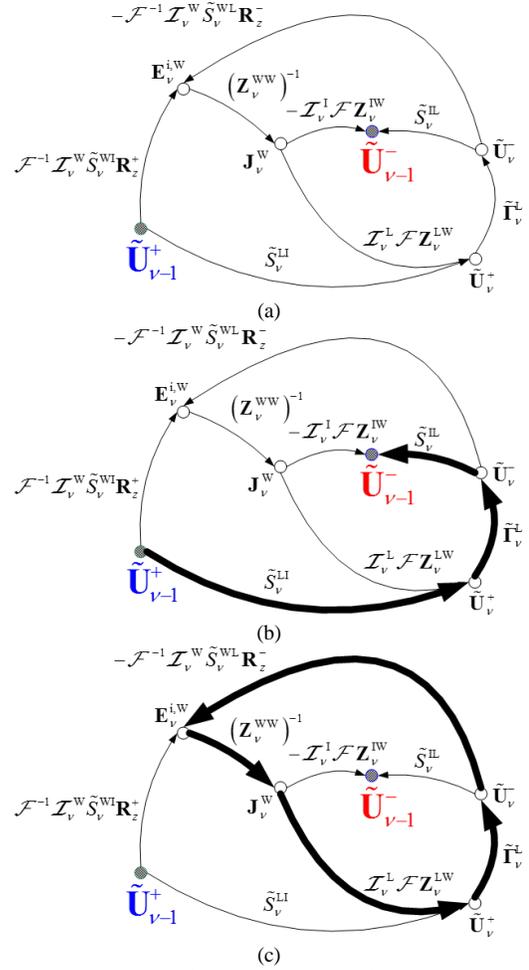

Fig. 4. A signal flow graph representation of the interaction in a given segment. (a) The signal flow graph representation of (36). (b) Direct reflection from the successive cross-sections. (c) Wall-Load cross-section loop.

The first term in (38) represents the direct reflection from the load interface, depicted by the thick arrows in Fig. 4(b) while the second term sums all the reflections from the segment's walls. This term is a multiplication of three factors: The first and the last account for reflection from- and incident on- the walls, respectively, while the factor in the middle represents the sum of multiple reflections between the walls and the load interface, depicted by the thick arrows in Fig. 4(c). Eq. (38) is the most compact representation of the reflection operator.

The CIO of the entire cavity is obtained now by applying (38) recursively, starting with $\tilde{\mathbf{\Gamma}}_L^L$, the load of the termination at the $v = L$ segment, and ending with $\tilde{\mathbf{\Gamma}}_0^L$ which is the Cavity's input reflection operator. If the termination is "simple", comprising only simple PEC walls, then its input reflection coefficient $\tilde{\Gamma}_L^L$ can be found via the following IE

$$\begin{pmatrix} 0 & -\mathcal{I}_{L+1}^I \mathcal{F} \mathbf{Z}_{L+1}^{IW} \\ -\mathcal{F}^{-1} \mathcal{I}_{L+1}^W \tilde{S}_{L+1}^{WI} \mathbf{R}_z^+ & \mathbf{Z}_{L+1}^{WW} \end{pmatrix} \begin{pmatrix} \tilde{\mathbf{U}}_L^+ \\ \mathbf{J}_{L+1}^W \end{pmatrix} = \begin{pmatrix} \tilde{\mathbf{U}}_L^- \\ 0 \end{pmatrix} \quad (39)$$

and $\tilde{\Gamma}_L^L$ is found by applying in (39) the Schur's complement procedure, obtaining

$$\tilde{\Gamma}_L^L = -\mathcal{I}_{L+1}^I \mathcal{F} \mathbf{Z}_{L+1}^{IW} \left( \mathbf{Z}_{L+1}^{WW} \right)^{-1} \mathcal{F}^{-1} \mathcal{I}_{L+1}^W \tilde{S}_{L+1}^{WI} \mathbf{R}_z^+ \quad (40)$$

If the termination is not simple, $\tilde{\Gamma}_L^L$ should be determined by other numerical scheme.

In order to solve the exterior IE, it is convenient to express the CIO by the impedance operator $\mathcal{Z}^A$ of (9). One way to compute $\mathcal{Z}^A$ is to use its direct spectral domain relation to the cavity reflection operator $\tilde{\Gamma}_0^L$. To that end we express (7) in the spectral domain

$$\tilde{\mathbf{M}}^A(\mathbf{k}_t) = \tilde{\mathcal{Z}}^A(\mathbf{k}_t, \mathbf{k}_t') \tilde{\mathbf{J}}^A(\mathbf{k}_t') \quad (41)$$

and

$$\tilde{\mathcal{Z}}^A(\mathbf{k}_t, \mathbf{k}_t') = \frac{1}{(2\pi)^2} \int d^2 x_t \int d^2 x_t' \, e^{j(\mathbf{k}_t \cdot \mathbf{x}_t - \mathbf{k}_t' \cdot \mathbf{x}_t')} \mathcal{Z}^A(\mathbf{x}_t, \mathbf{x}_t') \quad (42)$$

The desired relation between $\tilde{\mathcal{Z}}^A$ and $\tilde{\Gamma}_0^L$ is obtained by substituting (41) and (37) utilizing (34) into (33). The spatial operator $\mathcal{Z}^A$ is then calculated by performing the inverse Fourier transform of (42) for the spatial grid of the external IE.

An alternative approach to calculating $\mathcal{Z}^A$ is to solve the input segment $\nu = 1$ in a mixed spatial-spectral approach where the unknowns at the Input interface are described spatially using $(\mathbf{J}^A, \mathbf{M}^A)$ as in (7), while those at the Load interface are described spectrally using $\tilde{\mathbf{U}}_1^\pm$ which are related by $\tilde{\Gamma}_1^L$ which has already been calculated in the preceding step. The IE for that segment is given by

$$\begin{pmatrix} -\mathbf{Z}_1^{II} & -\mathbf{T}_1^{II} & \mathbf{Z}_1^{IW} & \mathcal{F}^{-1} S_1^{IL} \mathcal{I}_\nu^I \mathbf{R}_z^- \\ -\mathbf{Z}_1^{WI} & -\mathbf{T}_1^{WI} & \mathbf{Z}_1^{WW} & \mathcal{F}^{-1} S_1^{WL} \mathcal{I}_\nu^W \mathbf{R}_z^- \\ -\mathcal{I}_\nu^L \mathcal{F} \mathbf{Z}_1^{LI} & -\mathcal{I}_\nu^L \mathcal{F} \mathbf{T}_1^{LI} & \mathcal{I}_\nu^L \mathcal{F} \mathbf{Z}_1^{LW} & 0 \end{pmatrix} \cdot \begin{pmatrix} \mathbf{J}^A \\ \mathbf{M}^A \\ \mathbf{J}_1^W \\ \tilde{\mathbf{U}}_1^- \end{pmatrix} = \begin{pmatrix} 0 \\ 0 \\ \tilde{\mathbf{U}}_1^+ \end{pmatrix}. \quad (43)$$

The solution is given, in Appendix A, by (56) utilizing the operators defined in (66)-(71)

### C. Computation Complexity

The spectral formulation above has a number of favorable properties:
1) The segments are encapsulated in the sense that the inward and outward propagating waves are *a priori* separated, whereas separation in the spatial domain formulation requires integral operations.
2) Direct propagation between successive interfaces is a diagonal propagation operator. Cross-spectral coupling is due only to internal interactions with the segment's walls.
3) It can accommodate arbitrary cross sections.

The reduced computational complexity of the EDD stems from properties 1 and 2 above. Let the number of spectral unknowns at each coupling interface be $N^S$ and the number of spatial unknowns on the segment's walls be $N^W$. Thus $\tilde{\mathbf{U}}_\nu^\pm$ are $N^S$-vectors; the spectral propagators $\tilde{\mathbf{S}}_\nu^{IL}$ and $\tilde{\mathbf{S}}_\nu^{LI}$ are represented by a diagonal $N^S \times N^S$ matrices; $\tilde{\mathbf{S}}_\nu^{WI}$ and $\tilde{\mathbf{S}}_\nu^{WL}$ are $N^W \times N^S$ matrices; $\mathbf{Z}_\nu^{WW}$ is an $N^W \times N^W$ matrix representing the walls impedance matrix, and $\mathbf{Z}_\nu^{IW}$ and $\mathbf{Z}_\nu^{LW}$ are $N^S \times N^W$ matrices, representing the operators of computing the wall current contribution to spectral wave amplitudes on the respective coupling interfaces.

The computational cost in (38) stems from two contributors: the first term, the direct reflection path, and the second term, the wall-interfaces interactions. The spectral propagation operators in the first term of (38) are diagonal, hence the computational cost is $O((N^S)^2)$ In the second term, it is the wall-load loop that is the candidate to lead the computational complexity. We choose the segmentation step to be small compared to the transversal cross section such that $N^W \ll N^S$, hence the computational cost of inversion operations in (38) becomes negligible. Thus, the overall computational complexity of calculating the CIO via (38) is given by

$$CC^{EDD} \approx \frac{10}{3} L^{-2} (\hat{N}^C)^3 + 4L^{-1} N^S (\hat{N}^C)^2 + 6(N^S)^2 \hat{N}^C + L8(N^S)^2 \quad (44)$$

and the optimal number of segments is given by

$$L_{Opt}^{EDD} = \sqrt{(\hat{N}^C)^2 / (2N^S)} \quad (45)$$

where $L$, $N^A$, and $\hat{N}^C$ have been defined in (21), $N^S$ is the number of spectral elements where $O(N^S) = O(N^A)$. needed to store the preprocessed operators of a single segment is given by

$$Strg^{EDD} = (N^S)^2 + 2N^S + 4L^{-1} \hat{N}^C N^S + L^{-2} (\hat{N}^C)^2 \quad (46)$$

One readily discerns from (44) that for small $L$, $CC^{EDD}$ is governed by $\hat{N}^C$, while for large $L$ it is dominated by $L(N^S)^2$, direct interface reflections (1st term in (38)). Recalling that $CC^{SDD}$ in (21) is governed by $L(N^A)^3$ for large $L$, and comparing to $CC^{EDD}$ shows that the EDD is much more efficient then the SDD, as long as $O(N^S) = O(N^A)$. The optimal number of segments in the SDD and the EDD schemes, $L_{Opt}^{SDD}$ in (22) and $L_{Opt}^{EDD}$ in (45), suggests that in the SDD, the computation is inefficient for segment length is less than the aperture size, whereas in the EDD, the segmentation can go down to $O(\lambda)$. Thus the SDD does not benefit from inverting small operators, whereas the EDD does. The single segment storage requirements, $Strg^{EDD}$, in (46) is very similar to the $Strg^{SDD}$ in (23).

## IV. DISCRETIZATION

The EDD makes use of both spatial and spectral domains, hence, in order to numerically solve the OEC problem we discretize in both domains, wall currents in the spatial domain and segment coupling interfaces in the spectral domain. In the following section, we will discuss the discretization considerations in each domain.

### A. Spatial Discretization

For the spatial discretization, we utilize an LCN scheme where the surface currents, electric and magnetic, are represented, as well as for the testing points, by the quadrature rule. The key features of the LCN as discussed in [24]–[26] are: 1) LCN utilizes high-order basis functions with no cell-to-cell continuity condition which simplifies the formulation; 2) the continuity between successive elements is achieved by the high-order basis; 3) the use of high-order schemes accelerates the convergence with larger element sizes; 4) filling the matrices is extremely fast due to the use of sampling instead of integration. These features are exploited in our implementation, making it more natural to formulate and rapidly converging.

### B. Spectral Discretization

The main advantage of the EDD scheme stems from the spectral representation on the coupling interfaces. The spectral discretization is controlled by the spectral sampling rate (SSR) and the spectral bandwidth (SBW). The considerations for choosing these parameters are discussed below.

*1) Spectral Sampling Rate*

A sampled representation of the spectrum implies that the inner problem is extended periodically in the spatial transversal domain as shown in Fig. 5. The periodic-domain equivalence of (1), recalling that $\mathbf{E}^i = 0$ in the segments, is given by

$$0 = \mathbf{E}^P + \eta \mathcal{L}^P(\mathbf{J}) + \mathcal{K}^P(\mathbf{M}), \quad (47)$$

where $\mathbf{E}^P$ is the periodic extensions of $\mathbf{E}$ while the operators $\mathcal{L}^P$ and $\mathcal{K}^P$, the periodic extensions of $\mathcal{L}$ and $\mathcal{K}$, are the same expressions as in (3) but with the free space Green's function $G(\mathbf{r}, \mathbf{r}')$ replaced by the periodic Green's function given by

$$G^P(\mathbf{r}, \mathbf{r}'; D_x, D_y) = \sum_{m,n} G(\mathbf{r}, \mathbf{r}' + \hat{\mathbf{x}}mD_x + \hat{\mathbf{y}}nD_y). \quad (48)$$

where $D_x$ and $D_y$ denote the unit cell size in Cartesian coordinates. These are related to the discrete spectral domain wavenumbers via

$$\mathbf{k}_t^{p,q} = (k_x^p, k_y^q) = (2\pi p/D_x, 2\pi q/D_y) \quad (49)$$

and

$$k_z^{p,q} = \sqrt{k^2 - (k_t^{p,q})^2}, \qquad \text{Im}\{k_z^{p,q}\} \le 0. \quad (50)$$

The spectral representation of the surface integral operators in (47) is given now by sampling the operators in (26) and the transversal sources $(\tilde{\mathbf{J}}, \tilde{\mathbf{M}})$ on this discrete wavenumber lattice.

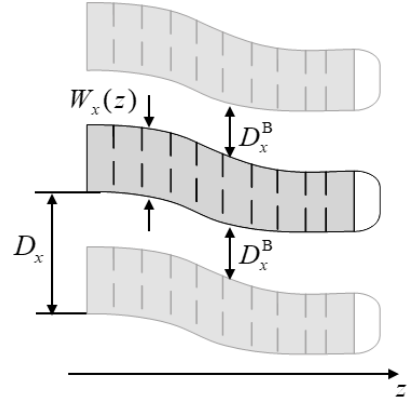

Fig. 5. OEC tangential width, $W_x(z)$, periodicity buffer, $D_x^B$, and periodicity length, $D_x$.

The unit cell dimensions, $D_x$ and $D_y$, are chosen such that there is no spatial overlap between the periodic replica of the segments. Fig. 5 schematizes the $x$-domain replica, where $W_x(z)$ denotes the cavity width along the $z$-axis, and $D_x^B$ is the periodicity buffer in the $x$ direction. In order to prevent a spatial aliasing, the periodicity lengths, $D_x$ and $D_y$ are chosen such that

$$D_{x,y} = \max_z (W_{x,y}(z)) + D_{x,y}^B. \quad (51)$$

Note that (48) does not converge and its Fourier sum, which is based on the spectral representation converges slowly as $z \to z'$, hence acceleration techniques, s. a. Ewald summation and Veysoglu transform [27]–[29], are used.

*2) Spectral Bandwidth*

The truncation rule of thumb is to have sufficient spectral bandwidth to accurately model the field discontinuity at the segment's wall in the sampling planes. In Section V, the dependence of the accuracy on the bandwidth is demonstrated.

*3) Spectral Singularity*

The spectral Green function has a branch point singularity at the transition from visible to evanescent regimes. The solution scheme is sensitive to the proximity to this spectral transition. The effect of spectral sampling near this transition point is explored numerically in Section V.

## V. NUMERICAL RESULTS

The performance of the EDD solution scheme is demonstrated on a 2D PEC configuration of an S-shaped OEC, embedded in an enclosing shell, as depicted in Fig. 6. We do not present the reduction of the 3D formulation above to the 2D case; the reader may find it the PhD dissertation of the first author, Analysis of Scattering from Large Open-Ended Cavities by Encapsulating Domain Decomposition. This paper presents the results only for the TM polarization; the results of the TE polarization will be presented briefly elsewhere.

The configuration dimensions (in units of $\lambda$) are: aperture width $W = 80$, inlet length in the z-direction (without the termination) $l_z = 200$, termination length 11, corners are 90° circular arcs with diameter of 10, inlet axis vertical shift $2a = 80$, and the enclosing shell extends 15 wavelength beyond the inlet bounding box. The S-Shape inlet center-line is given by $\mathbf{c}(t) = (a\cos(\pi t), tl_z)$, $t \in [0,1]$, and the cavity contour is described explicitly by the upper and lower walls, $\mathbf{c}^\pm(t)$, respectively, given by

$$c_x^\pm(t) = a\cos(\pi t) \pm \frac{W/2}{\sqrt{1+(\pi a l_z^{-1})^2 \sin^2(\pi t)}}, \quad (52)$$

and

$$c_z^\pm(t) = tl_z \pm \frac{(\pi a l_z^{-1} W/2)\sin(\pi t)}{\sqrt{1+(\pi a l_z^{-1})^2 \sin^2(\pi t)}}. \quad (53)$$

The geometry have been rendered to a 2nd order elements.

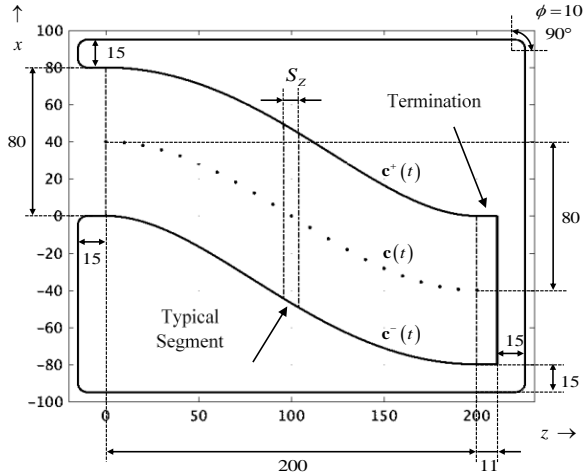

Fig. 6. Specifications of S-Shaped OEC embedded in an enclosing structure. All dimensions are in $\lambda$.

The EDD performance is demonstrated for 2D TM excitation and has been compared to several solution schemes: 1) reference solution, a brute-force solution of the whole problem, illustrated in Fig. 1(a), 2) $DD_0$ solution in which the problem has been decomposed into outer problem, enclosing shell, and inner problem, OEC, 3) SDD and EDD in which the inner problem has also been decomposed into segments $1.5\lambda$ in length. In the solution schemes, $DD_0$, SDD and EDD the spatial discretization has been maintained, 2nd order LCN with $\lambda/6$ element size, i.e. 18 samples per wavelength, which is equivalent to bandwidth of $18k_0$. As a reference, we utilized a solution with a higher accuracy, a 3rd order CFIE LCN with $\lambda/8$ element size has been utilized. The EDD with spectral bandwidth, $18k_0$, matches the spatial discretization, and $D_x^E$ of $\lambda/10$. The scattered field and RCS, presented in Fig. 7 for the problem defined in Fig. 6, have been calculated at infinity: Reference, $DD_0$, SDD, and EDD are depicted by the solid-blue, dashed-red, dotted-yellow and dash-dot-purple solid lines, respectively. $RCS/\lambda$ is displayed over a full azimuthal range (0°-180°) in Fig. 7(a), while Fig. 7(b) shows a zoom in on the region governed by the OEC, excluding the flash return from the enclosing shell walls, 95°-175°. The local error in the scattered electric field for each of the solution schemes, relative to the reference solution and normalized to the average scattered field of the reference solution, is given by

$$\Delta E^{\text{scat}} = 10 \log_{10} \left( \frac{\left| E^{\text{scat}}(\phi_n) - E^{\text{scat}}_{\text{Ref}}(\phi_n) \right|^2}{\frac{1}{N} \sum_{n=1}^{N} \left| E^{\text{scat}}_{\text{Ref}}(\phi_n) \right|^2} \right), \quad (54)$$

where $\phi_n, n = 1, \ldots, N$ is the angular sampling points. Fig. 7(c) presents the behavior of the error at 3601 equally spaced points over the full azimuthal range, 0°-180°. Here the sampling rate is about twice the Nyquist angular requirement for the configuration in Fig. 6. Also, $E^{\text{scat}}_{\text{Ref}}$ and $E^{\text{scat}}$ are the scattered electric field of the reference solution, and of the $DD_0$, SDD, or EDD schemes. The results, presented in Fig. 7, show a good agreement between all the solution schemes. In Figs. 8(a) and 8(b) the EDD RMS error demonstrated for various spectral bandwidth selections: $4.5k_0$, $9k_0$, $18k_0$, $36k_0$, and $72k_0$. The EDD sensitivity to periodic buffer, $D_x^B$, defined in Fig. 5, is given in Figs. 8(a) and 8(b). The accuracy in Figs. 8(a) and 8(b) is measured by an RMS error estimator, given by

$$E^{\text{scat}}_{\text{NE}} = 10 \log_{10} \left( \frac{\sum_{n=1}^{N} \left| E^{\text{scat}}(\phi_n) - E^{\text{scat}}_{\text{Ref}}(\phi_n) \right|^2}{\sum_{n=1}^{N} \left| E^{\text{scat}}_{\text{Ref}}(\phi_n) \right|^2} \right) \quad (55)$$

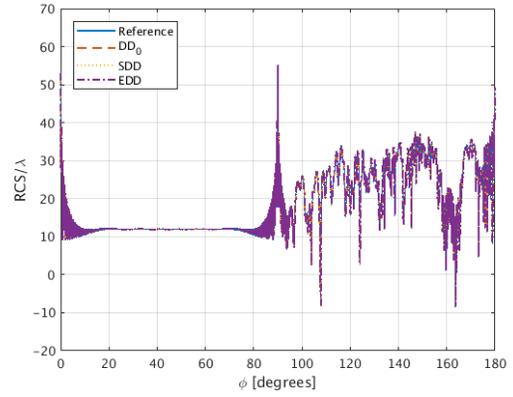

(a)

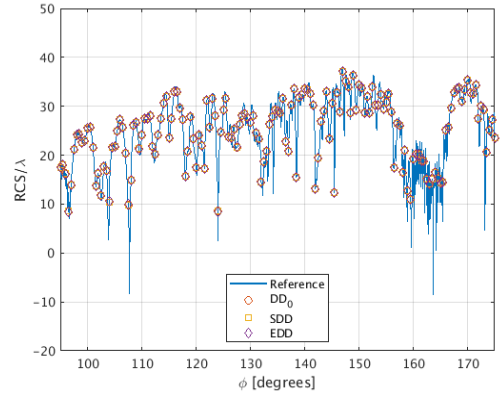

(b)

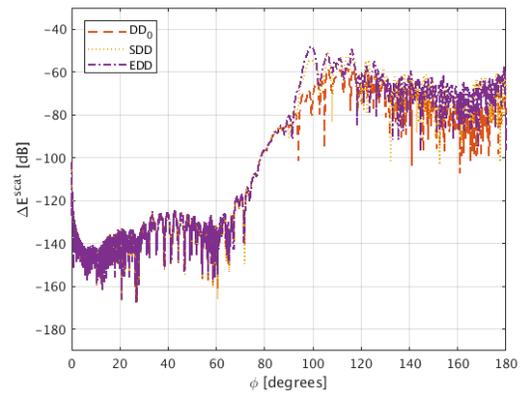

(c)

Fig. 7. S-Shape OEC RCS plot of four solution schemes. Reference solution in solid-blue, $DD_0$ in dash-red, spatial SDD in dot-yellow, and EDD in dot-dash-purple. In (a), a full azimuth span, in (b), zoom on azimuth 95°-175°, region that governed by the OEC, and in (c) the error between the scattered electric field of each solution and the reference solution, divided by maximum scattered field.

The region of interest 95°-175° has been selected to exclude the flash return from the enclosing shell walls. As shown in Figs. 8(a) and 8(b), the periodic buffer, $D_x^B$, has the following constrains: 1) There is a breakdown in the accuracy for $D_x^B$ values that are lower than $\lambda/10$, 2) as we get closer to the spectral branch point, $k_x^p \to k_0$, the solution scheme accuracy deteriorates. The measured computational speedup for the overall OEC is presented in Fig. 9. The speedup has been measured for the following segment steps, $S_Z$ : 1, 2, 4, 8, 15.4, 28.6, 50 and 100 $\lambda$. Fig. 9 shows that as the ratio between aperture width and segment length, increases, the EDD become more efficient than the SDD and that they are spectral bandwidth dependent. Using a bandwidth wider than twice the spatial bandwidth doesn't improve the accuracy and will result in an inefficient scheme.

## VI. Conclusions and Summary

In this paper, the EDD scheme have been presented in detail. The efficiency of the EDD has been compared to the SDD and $DD_0$, with respect to the following efficiency criteria: 1) accuracy of calculations for an arbitrarily shaped OEC, 2) computational complexity, and 3) storage requirements.
1) **Accuracy of calculating an arbitrarily shaped OEC:** The results show that the EDD agrees with the spatial solution schemes and the error trends are similar as well.
2) **Computational Speedup:** The speedup of the EDD depends on two parameters: spectral bandwidth and

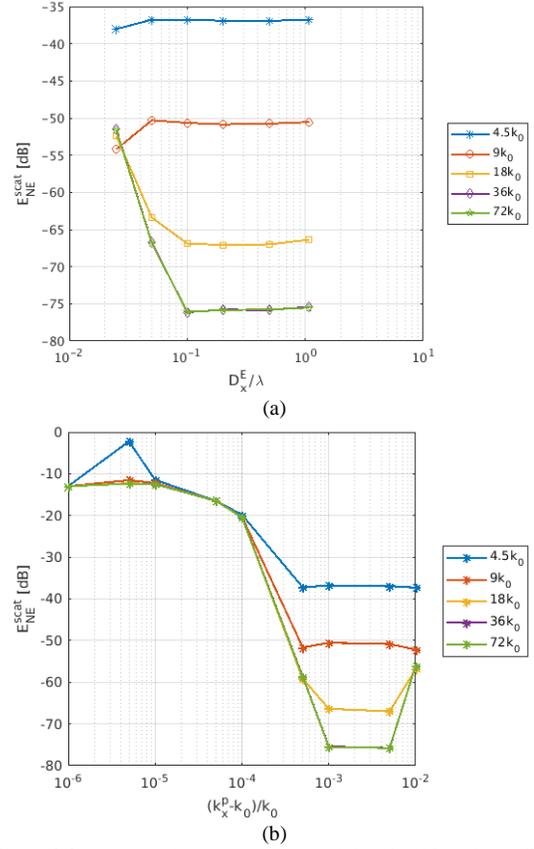

Fig. 8. EDD accuracy for various spectral bandwidth values. Spectral bandwidth of $4.5k_0$ in blue-asterisk, $9k_0$ in red-circle, $18k_0$ in yellow-square, $36k_0$ in purple-diamond, and $72k_0$ in green-pentagram. In (a) RMS of the scattered electric field as a function of periodic extension, $D_x^B$, and in (b) as a function of proximity to the spectral singularity.

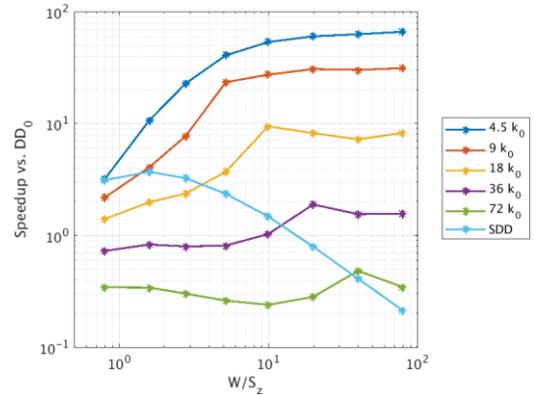

Fig. 9. EDD and SDD vs. $DD_0$ measured speedup as function of representing aperture width (average), W, to segment length, $S_Z$, ratio for various spectral bandwidth values: Spectral bandwidth of $4.5k_0$ in blue-asterisk, $9k_0$ in red-circle, $18k_0$ in yellow-square, $36k_0$ in purple-diamond, and $72k_0$ in green-pentagram.

aperture width, W, to segment length, $S_z$, ratio, for a given cavity size. The straightforward spectral bandwidth choice is to match the spatial sampling rate (in our case $18k_0$), which result in an accurate solution and a minimal speedup of 5. A twice wider spectral bandwidth will result in a better accuracy on the expense of lower speedup for the same aperture to segment length ratio, but the accuracy doesn't get better for a wider spectral bandwidth. If lower accuracy is acceptable, a narrower spectral bandwidth can be used which accelerates the computation even further.

3) **Storage requirements:** The EDD requires almost the same storage as the one of the SDD. The reflection operators are slightly bigger than the ones in the SDD, whereas the aperture to aperture propagation is represented by diagonal operators, hence stored as one-dimensional vectors. Compared to solving the OEC utilizing $DD_0$, the EDD and the SDD storage requirements are significantly reduced.

## APPENDIX A. IMPEDANCE INPUT OPERATOR OF AN OEC SEGMENT

The explicit expression of the Impedance IO of an OEC segment can be found by substituting (19) into (18) and solving for the relation between $\mathbf{M}_{\nu-1}^L$ and $\mathbf{J}_{\nu-1}^L$. The resulting expression:

$$\mathcal{Z}_{\nu-1}^L = -\left(\overline{\mathbf{T}}_\nu^{II} - \overline{\mathbf{Z}}_\nu^{IW}\left(\overline{\mathbf{Z}}_\nu^{WW}\right)^{-1}\overline{\mathbf{T}}_\nu^{WI}\right)^{-1} \left(\overline{\mathbf{Z}}_\nu^{II} - \overline{\mathbf{Z}}_\nu^{IW}\left(\overline{\mathbf{Z}}_\nu^{WW}\right)^{-1}\overline{\mathbf{Z}}_\nu^{WI}\right) \quad (56)$$

is a generalization of (9), where the bar on top of the operators denotes generalized operators defined as follows:

$$\overline{\mathbf{Z}}_\nu^{WW} = \mathbf{Z}_\nu^{WW} - \overline{\mathbf{Z}}_\nu^{WL}\left(\overline{\mathbf{Z}}_\nu^{LL}\right)^{-1}\mathbf{Z}_\nu^{LW}, \quad (57)$$

$$\overline{\mathbf{T}}_\nu^{II} = \mathbf{T}_\nu^{II} - \overline{\mathbf{Z}}_\nu^{IL}\left(\overline{\mathbf{Z}}_\nu^{LL}\right)^{-1}\mathbf{T}_\nu^{LI}, \quad (58)$$

$$\overline{\mathbf{T}}_\nu^{WI} = \mathbf{T}_\nu^{WI} - \overline{\mathbf{Z}}_\nu^{WL}\left(\overline{\mathbf{Z}}_\nu^{LL}\right)^{-1}\mathbf{T}_\nu^{LI}, \quad (59)$$

$$\overline{\mathbf{Z}}_\nu^{II} = \mathbf{Z}_\nu^{II} - \overline{\mathbf{Z}}_\nu^{IL}\left(\overline{\mathbf{Z}}_\nu^{LL}\right)^{-1}\mathbf{Z}_\nu^{LI}, \quad (60)$$

$$\overline{\mathbf{Z}}_\nu^{IW} = \mathbf{Z}_\nu^{IW} - \overline{\mathbf{Z}}_\nu^{IL}\left(\overline{\mathbf{Z}}_\nu^{LL}\right)^{-1}\mathbf{Z}_\nu^{LW}, \quad (61)$$

and

$$\overline{\mathbf{Z}}_\nu^{WI} = \mathbf{Z}_\nu^{WI} - \overline{\mathbf{Z}}_\nu^{WL}\left(\overline{\mathbf{Z}}_\nu^{LL}\right)^{-1}\mathbf{Z}_\nu^{LI}. \quad (62)$$

Here $\overline{\mathbf{Z}}_\nu^{LL}$, $\overline{\mathbf{Z}}_\nu^{WL}$, and $\overline{\mathbf{Z}}_\nu^{IL}$ are given by

$$\overline{\mathbf{Z}}_\nu^{LL} = \mathbf{Z}_\nu^{LL} + \mathbf{T}_\nu^{LL}\mathcal{Z}_\nu^L \quad (63)$$

$$\overline{\mathbf{Z}}_\nu^{WL} = \mathbf{Z}_\nu^{WL} + \mathbf{T}_\nu^{WL}\mathcal{Z}_\nu^L \quad (64)$$

$$\overline{\mathbf{Z}}_\nu^{IL} = \mathbf{Z}_\nu^{IL} + \mathbf{T}_\nu^{IL}\mathcal{Z}_\nu^L \quad (65)$$

with $\mathcal{Z}_\nu^L$ defined in (19). A generalized operator represents all the interaction between source and observation regions, including indirect interactions between these regions via the load. For example, the operator in (57) represents wall-to-wall interactions where $\mathbf{Z}_\nu^{WW}$ is the direct self-interaction of the wall, $\mathbf{Z}_\nu^{LW}$ represents the induced electric field load region which is induced by $\mathbf{J}_\nu^W$, the equivalence electric current on the wall region. The operator $\overline{\mathbf{Z}}_\nu^{LL}$, given in (63), is the sum of the direct load-to-load interaction and all the successive inlet segments which are represented by (19). In a similar manner, the operator $\overline{\mathbf{Z}}_\nu^{WL}$, given in (64), is the sum of the direct load-to-wall interaction and all the successive inlet segments to the wall region of the given OEC segment, which are represented by (19). Similarly, the operator $\overline{\mathbf{Z}}_\nu^{IL}$, given in (65), is the sum of the direct load-to-input interaction and all the successive inlet segments to the input region of the current OEC segment, which are represented by (19).

## APPENDIX B. SPECTRAL TO SPATIAL TRANSFORMATION

The above transforms the spectral plane wave on the load interface to a spatial current distribution on the input interface, and by that, enables us, easily, to connect the CIO with outer problem. Basically, the explicit expression of (43) is identical to (56) except for the generalized operators where the generalized operators are

$$\overline{\mathbf{T}}^{II} = \mathbf{T}_1^{II} - \mathcal{F}^{-1} S_1^{IL} \mathcal{I}_\nu^I \mathbf{R}_z^- \mathbf{\Gamma}_1^L \mathcal{I}_\nu^L \mathcal{F} \mathbf{T}_1^{LI}. \quad (66)$$

$$\overline{\mathbf{T}}^{WI} = \mathbf{T}_1^{WI} - \mathcal{F}^{-1} S_1^{WL} \mathcal{I}_\nu^W \mathbf{R}_z^- \mathbf{\Gamma}_1^L \mathcal{I}_\nu^L \mathcal{F} \mathbf{T}_1^{LI}. \quad (67)$$

$$\overline{\mathbf{Z}}^{II} = \mathbf{Z}_1^{II} - \mathcal{F}^{-1} S_1^{IL} \mathcal{I}_\nu^I \mathbf{R}_z^- \mathbf{\Gamma}_1^L \mathcal{I}_\nu^L \mathcal{F} \mathbf{Z}_1^{LI}. \quad (68)$$

$$\overline{\mathbf{Z}}^{WI} = \mathbf{Z}_1^{WI} - \mathcal{F}^{-1} S_1^{WL} \mathcal{I}_\nu^W \mathbf{R}_z^- \mathbf{\Gamma}_1^L \mathcal{I}_\nu^L \mathcal{F} \mathbf{Z}_1^{LI}. \quad (69)$$

$$\overline{\mathbf{Z}}^{IW} = \mathbf{Z}_1^{IW} - \mathcal{F}^{-1} S_1^{IL} \mathcal{I}_\nu^I \mathbf{R}_z^- \mathbf{\Gamma}_1^L \mathcal{I}_\nu^L \mathcal{F} \mathbf{Z}_1^{LW}. \quad (70)$$

$$\overline{\mathbf{Z}}^{WW} = \mathbf{Z}_1^{WW} - \mathcal{F}^{-1} S_1^{WL} \mathcal{I}_\nu^W \mathbf{R}_z^- \mathbf{\Gamma}_1^L \mathcal{I}_\nu^L \mathcal{F} \mathbf{Z}_1^{LW}. \quad (71)$$